\documentclass[english,aps,manuscript,showpacs,amssymb,amsfonts]{revtex4}
\usepackage[T1]{fontenc}
\usepackage[latin1]{inputenc}
\usepackage{amsmath}
\usepackage{babel}
\usepackage{graphics}

\makeatletter

\providecommand{\LyX}{L\kern-.1667em\lower.25em\hbox{Y}\kern-.125emX\@}

\makeatother

\begin{document}

\preprint{DAMTP-2003-95}

\preprint{hep-th/0310067}


\title{Gravity \textit{{\`a} la} Born-Infeld}

\author{Mattias N.R. Wohlfarth}

\email{M.N.R.Wohlfarth@damtp.cam.ac.uk}

\affiliation{Department of Applied Mathematics and Theoretical Physics, Centre for Mathematical
Sciences, University of Cambridge, Wilberforce Road, Cambridge CB3 0WA, United
Kingdom}

\begin{abstract}
A simple technique for the construction of gravity theories in Born-Infeld style
is presented, and the properties of some of these novel theories are investigated.
They regularize the positive energy Schwarzschild singularity, and a large class
of models allows for the cancellation of ghosts. The possible correspondence
to low energy string theory is discussed. By including curvature
corrections to all orders in \( \alpha '\), the new theories nicely
illustrate a mechanism that string theory might use to regularize
gravitational singularities. 
\end{abstract}
\maketitle

\section{Introduction}

The problem of the diverging Coulomb field and self-energy of point particles
in Maxwell's theory of electrodynamics led Born and Infeld in the 1930s to the
construction of a non-linear extension of this theory \cite{BoIn1934} in which
a regularization is achieved because the field strength tensor \( F_{ab} \)
appears to all orders. A particular appeal of Born-Infeld electrodynamics is
that its action can be expressed in the compact determinantal form
\begin{equation}
\int \sqrt{-\det (\eta _{ab}+\lambda F_{ab})}
\end{equation}
and, more importantly, it stems from the fact that the Dirac-Born-Infeld action
resurfaces in string theory as the low energy effective action \cite{FrTs1985}
on the worldvolume of \( D \)-branes, by capturing the degrees of freedom of
open strings in an abelian background gauge field. The Born-Infeld parameter
\( \lambda  \) is then related to the inverse tension of fundamental strings,
by \( \lambda =2\pi \alpha'  \).

Einstein's general relativity is the fundamental building block for the gravity
sector of superstring or M-theory which consistently unify gravity and the standard
model of elementary particle physics. In supersymmetrized form, general relativity
appears as the low energy effective theory (in ten or eleven dimensions). That
it is plagued, like Maxwell electrodynamics, by singularities in the gravitational
field of point masses, thus can be considered a major problem. Of course, string/M-theory
predict corrections to general relativity at higher orders in the curvature
(corresponding to higher orders in \( \alpha ' \)), but no way has
been found yet to resum these corrections, as in the Born-Infeld case, so to incorporate
arbitrary high powers of the curvature and to solve the problems inherent in
Einstein gravity.

Inspired by Born-Infeld electrodynamics, several attempts have been made in
the construction of gravity analogues. Deser and Gibbons \cite{DeGi1998} formulate
three obvious criteria that such a theory should satisfy.

\begin{enumerate}
\item Freedom of ghosts.
\item Regularization of singularities. 
\item Supersymmetrizability.
\end{enumerate}
Some comments on these points are in order. 

\textit{ad 1}. Responsible for the propagation of ghosts, in a perturbative
expansion around the Minkowski vacuum spacetime, are the terms in the action
which are quadratic in the curvature. The only exemption from this rule is the
quadratic Gauss-Bonnet combination
\begin{equation}
\label{eq. gbcomb}
\sqrt{-g}\, (R^{2}-4R_{ab}^{2}+R_{abcd}^{2})
\end{equation}
which is a total derivative in four dimensions and for which, in higher dimensions,
the quadratic terms in the perturbative expansion still cancel up to total derivatives
\cite{Zwi1985}. 

\textit{ad 2}. In general, not all types of singularities can be removed, and
not all of them should be removed, as pointed out by Horowitz and Myers \cite{HoMy1995}.
They argue that certain gravitational plane waves are solutions to any general
theory whose Lagrangian is a function of the metric, the curvature and its covariant
derivatives. This is due to the fact that all curvature scalars of these spacetimes
vanish. But the two polarizations of a plane wave can diverge, which leads to
unbounded tidal forces. Hence, all extensions of general relativity will have
solutions with null singularities. Furthermore, negative energy solutions of
general relativity, e.g., the Schwarzschild solution with negative mass parameter,
should not be regularized, since this would imply an instability of the Minkowski
vacuum. The singularities of negative energy solutions serve as a useful criterion
of unphysicality. 

\textit{ad 3}. Supersymmetrizability is a very stringent requirement and is
presumably natural if the gravity theory were implied by higher-dimensional
string/M-theory.

It is illuminating to take a closer look at previous attempts to construct gravities
in Born-Infeld style. The discussion of \cite{DeGi1998}, for instance, considers
a Lagrangian of the form \( (-\det (c_{1}g_{ab}+c_{2}R_{ab}+c_{3}X_{ab}))^{1/2} \)
where the tensor \( X_{ab} \) contains terms of second or higher order in the
curvature and the \( c_{i} \) are arbitrary constants. The simplest of these
models has \( c_{1}=1 \) and \( c_{3}=0 \). It is, of course, not ghost-free
because terms from \( X_{ab} \) are needed to balance those arising in second
order of the curvature expansion from \( R_{ab} \). This model does not improve
general relativity in the sense that it has the standard Schwarzschild (-de
Sitter) solution. Note that this can be explained intuitively. The Schwarzschild
solution is Ricci flat, and hence, its singularity can be recognized only from
the second order curvature invariant \( R_{abcd}^{2} \), but not from \( R^{2} \)
and \( R_{ab}^{2} \). But this invariant does not occur in the curvature expansion
of the above model, so that it may be unbounded in solutions. A theory that
places an upper bound on \( R_{abcd}^{2} \) is introduced in a rather \textit{ad
hoc} way in \cite{Fei1998}, essentially by adding \( c_{1}(1-c_{2}R_{abcd}^{2})^{1/2} \)
to the Einstein-Hilbert Lagrangian. The author of this paper finds non-singular
black hole solutions, which underlines the importance of using the full Riemann
tensor in order to regularize gravitational singularities. Supersymmetric constructions,
which will not be further discussed here, can be found in \cite{GaKe2001}.

Making use of these insights, in this article, a method is presented to construct
a large, and novel, class of gravity theories in Born-Infeld style. Section
\ref{sec. basic model} introduces the basic model and discusses some of its
properties. In particular, it is found that the positive mass Schwarzschild
singularity may be removed, whereas the negative energy solution remains singular
as is necessary for a stable Minkowski vacuum. Section \ref{sec. cancel ghosts}
considers generalizations of the basic model that allow for the cancellation
of ghosts. These generalized models are compared and matched to the low energy
effective gravity action following from string theory, in section \ref{sec. string theory}.
The article concludes with a discussion in section \ref{sec. discussion}.

\section{The basic model\label{sec. basic model}}

\subsection{Action and equations of motion}

Start out with the following two basic requirements. In order to be as close
in spirit as possible to the structure of Born-Infeld electrodynamics, try to
find a gravity action, on a \( d \)-dimensional pseudo-Riemannian manifold
with metric \( g_{ab} \), which can be written in a nice determinantal form,
and incorporate the full Riemann tensor in order to enable a possible regularization
of the singularities. Both requirements can be satisfied by using the symmetries
of the Riemann tensor calculated from the metric connection. It is antisymmetric
in two pairs of indices, i.e., it satisfies \( R_{abcd}=R_{[ab]cd}=R_{ab[cd]} \).
Introduce the symmetric tensor
\begin{equation}
R_{AB}\equiv R_{[a_{1}a_{2}][b_{1}b_{2}]}
\end{equation}
whose capital indices take \( d(d-1)/2 \) values that can be imagined as ordered
pairs of standard indices. Along with this definition, introduce a new metric
and Kronecker delta obeying the same symmetries,\begin{subequations}
\begin{eqnarray}
g_{AB} & \equiv  & g_{a_{1}b_{1}}g_{a_{2}b_{2}}-g_{a_{2}b_{1}}g_{a_{1}b_{2}}\, ,\\
\delta ^{A}_{B} & \equiv  & \delta ^{a_{1}}_{b_{1}}\delta ^{a_{2}}_{b_{2}}-\delta ^{a_{2}}_{b_{1}}\delta ^{a_{1}}_{b_{2}}\, .
\end{eqnarray}
\end{subequations}These tensors can be used to upper and lower, or replace,
capital indices as usual. For the metric, a useful determinant formula holds,
\begin{equation}
\det g_{AB}=(\det g_{ab})^{d-1}\, .
\end{equation}

This suggests the following basic model of a Born-Infeld type gravity, valid
in an arbitrary dimension \( d \). Consider the action
\begin{equation}
\int \left( -\det (g_{AB}+\lambda R_{AB})\right) ^{\zeta }=\int \sqrt{-g}\left( \det (\delta ^{A}_{B}+\lambda R^{A}_{\, \, B})\right) ^{\zeta }
\end{equation}
with a parameter \( \lambda  \) of the dimension length squared. The equality
holds for an exponent \( \zeta =\frac{1}{2(d-1)} \). Note, however, that one
might start out from the action on the right hand side which allows for different
\( \zeta  \). The only restriction appears to be a fractional \( \zeta  \)
because a possible regularization of singularities is only expected from the
appearance of all curvature orders in the action. A curvature expansion of the
Lagrangian is equivalent to an expansion around \( \lambda =0 \) and, being
a special case of equation (\ref{eq. curv exp}) with \( M^{A}_{\, \, B}=R^{A}_{\, \, B} \)
and \( N^{A}_{\, \, B}=0 \), gives
\begin{equation}
1+\frac{1}{2}\zeta \lambda \left( R+\frac{1}{4}\zeta \lambda R^{2}-\frac{1}{4}\lambda R_{abcd}^{2}\right) +\mathcal{O}(\lambda ^{3})
\end{equation}
for the lowest orders. This shows that it is necessary to subtract the cosmological
constant that is implicit above, in order to obtain the Einstein-Hilbert Lagrangian
in the limit \( \lambda \rightarrow 0 \). The amended action considered in
the following is
\begin{equation}
\label{eq. basic action}
\int \sqrt{-g}\left[ \left( \det (\delta ^{A}_{B}+\lambda R^{A}_{\, \, B})\right) ^{\zeta }-1\right] .
\end{equation}
For small \( \lambda  \) this theory inherits all experimental tests of general
relativity. The curvature expansion, however, also shows that a quantization
of this classical theory would include ghost modes because the quadratic curvature
terms do not appear in the Gauss-Bonnet combination; see (\ref{eq. gbcomb})
in the introduction. The cancellation of the ghosts will be discussed in the
following section \ref{sec. cancel ghosts}. 

For the moment, it is useful to push on and to investigate some properties of
this basic model which are carried over, at least qualitatively, to the more
complicated models. The equations of motion are derived by varying the action
(\ref{eq. basic action}) with respect to the standard spacetime metric \( g_{ab} \).
They read
\begin{eqnarray}
 & \left( \det (\delta +\lambda R)\right) ^{\zeta }\left[ g^{cd}+\frac{1}{2}\zeta \lambda \, (\delta +\lambda R)^{-1\, B}_{\quad \, \, \, \, a_{1}a_{2}}\left( g^{c[a_{1}}R^{a_{2}]d}_{\quad \, B}+g^{d[a_{1}}R^{a_{2}]c}_{\quad \, B}\right) \right] -g^{cd} & \nonumber \\
 & +\, \frac{1}{2}\zeta \lambda \, \delta ^{c}_{[b_{1}}g^{d[a_{1}}\left( \nabla ^{a_{2}]}\nabla _{b_{2}]}+\nabla _{b_{2}]}\nabla ^{a_{2}]}\right) \left[ \left( \det (\delta +\lambda R)\right) ^{\zeta }(\delta +\lambda R)^{-1\, b_{1}b_{2}}_{\quad \quad \, \, a_{1}a_{2}}\right] =\, 0\, . & 
\end{eqnarray}
Terms in \( \mathcal{O}(1) \) cancel in the limit \( \lambda \rightarrow 0 \).
This is enforced by the subtraction of the cosmological constant in the action,
which is responsible for the subtraction of the term \( -g^{cd} \) in the equations
of motion. The second line is of \( \mathcal{O}(\lambda ^{2}) \) and vanishes.
So, in \( \mathcal{O}(\lambda ) \), Einstein's vacuum equations \( R^{cd}-\frac{1}{2}Rg^{cd}=0 \)
remain from the first line, as they should.

\subsection{Schwarzschild without singularities}\label{sec. sing}

Now consider the analogue of the spherically symmetric Schwarzschild solution
of general relativity. It turns out to be easier, than using the equations of
motion given above, to derive an effective action for the metric ansatz
\begin{equation}
g_{ab}=\textrm{diag}(-A(r),\, B(r),\, r^{2},\, r^{2}\sin ^{2}\theta )
\end{equation}
where the standard Schwarzschild coordinates are given by \( \{t,\,
r,\, \theta ,\, \phi \} \).
Such an approach usually is only valid in very special situations,
where it can be shown that the symmetry reduction by means of a
specific ansatz and the variational principle, used to derive the
equations of motion, commute \cite{Pal1979}. This is true for the
spherically symmetric ansatz and has been widely used in the
literature, e.g., in \cite{FeTo2002,DeTe2003}. The effective action is obtained to be
\begin{eqnarray}
 &  & \int dr\, rAB\left[ \frac{1}{2}A^{-7/6}B^{-2}\left( (\lambda +B(r^{2}-\lambda ))(2rAB+\lambda A')^{2}(2rB^{2}-\lambda B')^{2}\right. \right. \nonumber \\
 &  & \qquad \qquad \qquad \qquad \qquad \left. \left. \times (4A^{2}B^{2}-\lambda (AA'B'+A'^{2}B-2AA''B))\right) ^{1/6}-1\right] 
\end{eqnarray}
in four dimensions and with the exponent \( \zeta =\frac{1}{2(d-1)} \). (The
results of this section are expected to hold qualitatively for higher \( d \)
and different fractional exponent \( \zeta  \) as well. Five dimensions and,
for \( \zeta  \), various fractional values between zero and one have been
checked numerically.) 

Inspired by the general relativity solution, simplify
by setting \( B=1/A \). Now again, it is not immediately clear whether
the effective action approach stays valid under this additional assumption, and it is also not obvious
which symmetry might be responsible for that. The specific
solutions which will be derived below, however, can be checked to
satisfy also the separate equations of motion for \( A \) and \(
B \) following from the above action. So here, \( B=1/A \) is
admissible (this is, for instance, not the case for the
Einstein-Hilbert action). The term associated to the subtraction of the
minus one becomes independent of the function \( A \) and can be dropped. Furthermore,
the action can be rewritten compactly in terms of an auxiliary function
\begin{equation}
\label{eq. Ddef}
D(r)=r^{2}+\lambda (A(r)-1)
\end{equation}
as
\begin{equation}
\int dr\, r\left( DD'^{4}D''\right) ^{1/6}.
\end{equation}
The equation of motion, finally, becomes a fourth order ordinary differential
equation for the function \( D(r) \),
\begin{eqnarray}
 & 5rD'^{4}D''^{2}-2DD'^{2}D''\left[ 13rD''^{2}+D'(6D''-5rD''')\right] +D^{2}\left[ -40rD''^{4}\right.  & \nonumber \\
 & \left. +\, 8D'D''^{2}(12D''+5rD''')+5D'^{2}\left( -11rD'''^{2}+6D''(2D'''+rD'''')\right) \right] \, =\, 0\, . & 
\end{eqnarray}
The `trivial' solutions of this equation, analytic in \( r \) around
\( r=0 \), are \( c_{1}r^{3},\, c_{1}r^{2} \)
and \( c_{1}r+c_{2} \) for arbitrary constants \( c_{i} \). Of these solutions,
\( D(r)=r^{2} \) gives Minkowski space where \( A(r)=1 \). 

In order to calculate the spherically symmetric gravitational field, modified
with respect to general relativity, and to obtain Minkowski space asymptotically,
the boundary condition \( A(r)\rightarrow 1 \) as \( r\rightarrow \infty  \)
has to be satisfied. This suggests a solution for \( D(r) \) in terms of a
power series, essentially in the inverse radius \( r^{-1} \),
\begin{equation}
D(r)=r^{2}(1+\sum _{n=1}^{\infty }a_{3n}r^{-3n})\, ,
\end{equation}
where it turns out that only every third term contributes. A substitution of
this expansion into the equation of motion allows solving sequentially for the
unknown coefficients \( a_{3n} \) that are found to have the form \( a_{3n}=b_{3n}a_{3}^{n} \)
for \( n>1 \) and a set of positive numbers \( b_{3n} \), e.g., \( b_{6}=\frac{11}{94},\, b_{9}=\frac{1163}{19317}\, ...\,  \).
From the Schwarzschild boundary condition, \( A(r)\sim 1-\frac{2m}{r} \) as
\( r \) approaches infinity, and from the definition of \( D(r) \) in (\ref{eq. Ddef})
follows a relation of the coefficient \( a_{3} \) to the Schwarzschild mass
parameter \( m \), 
\begin{equation}
a_{3}=-2m\lambda \, .
\end{equation}

Look at the analogue of the positive energy Schwarzschild solution
first, i.e., set \( m>0 \) and \( \lambda>0 \).
 The convergence of the power series solution breaks down for \(
r^3<r_0^{3}\equiv |a_{3}| \).
A numerical solution of the differential equation, for negative \( a_{3} \), can be obtained up to
this point which turns out to be special. At \( r_0 \), the function
\( D \) vanishes, but \( D'(r_0)>0 \). The equation of motion is
satisfied by \( D''(r_0)=0 \). The second derivative at this point, however,
tends to zero with an infinite positive slope and an infinite negative
curvature. So the solution for \( D \) becomes singular at \( r_0 \). In
principle, the numerical integration of the equation of motion
could be carried on to smaller values \( r<r_0 \), but this would
necessitate an exact knowledge of the type of singularity involved. 

\begin{figure}[ht]
{\par\centering \resizebox*{0.65\columnwidth}{!}{\includegraphics{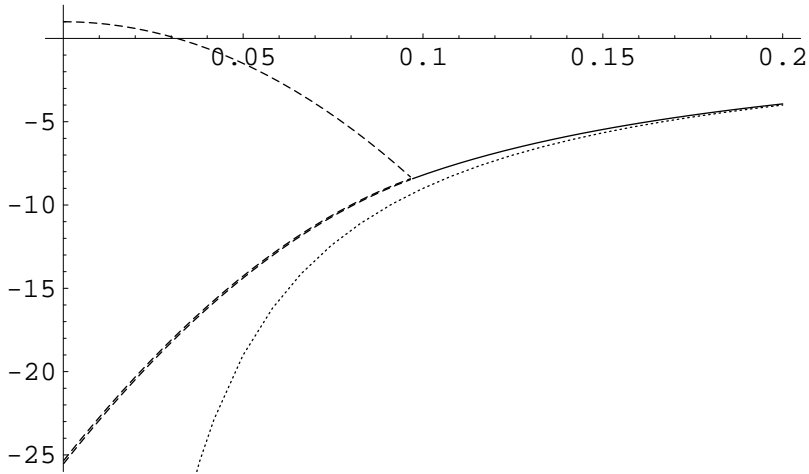}} \par}
\caption{\label{fig. Alimit}\textit{The solid line shows the numerical solution obtained
for the Schwarzschild function \protect\( A(r)\protect \) for \(
r>r_0=0.1\protect \). The possible matching solutions for \(
r<r_0\protect \) are plotted in wide dashes. The upper curve
represents an identically vanishing \( D(r)\protect \), the two, very
similar, lower curves
are obtained, respectively, from a linear \( D(r)\protect \) and from numerical
integration, as explained in the text.
The curve in close dashes presents the standard Schwarzschild solution
\protect\( A(r)=1-\frac{2m}{r}\protect \) of general relativity with its horizon
fixed at the value \protect\( r=1\protect \). }}
\end{figure}

Lacking such knowledge about the singularity of the solution at \(
r_0 \), it is still interesting to ask which solutions on the inner
domain might match to the outer one. A way to integrate further
inwards is to simply assume a pole in the third derivative of \( D \)
and to supply new initial conditions at some value of \( r \) slightly
smaller than \( r_0 \). Another possibility is to look for solutions
to the equation of motion which are analytic in \( r \) at \( r=0 \)
and obey the very weak condition that their function value match the
outer solution at \( r_0 \). The latter case is extremely restrictive,
and there are only two
possibilities. The first is the linear function \( D(r)=D'(r_0) (r-r_0) \), and the
second is an identically vanishing \( D(r)\equiv 0 \). In any case, it is
impossible to smoothly match all derivatives at \( r_0 \). The outer
solution and the possible inner solutions are shown in figure
\ref{fig. Alimit}; the numerically continued solution approximates the
linear \( D(r) \) to a high degree of accuracy.
Note that all possible inner solutions regularize the metric function
\( A(r) \) at \( r=0 \). The curvature singularities, however, are only
regularized by the inner solution with vanishing \( D(r) \). In this
case, one finds \( R=-\frac{6}{\lambda},\,
R_{ab}^2=\frac{36}{\lambda^2} \) and \( R_{abcd}^2=\frac{6}{\lambda^2}
\) throughout the inner domain.   

The completely regularized solution is very interesting for the
following reason. Vanishing \(
D(r)\equiv 0 \) corresponds to a Schwarzschild function \(
A(r)=1-r^2/\lambda \) on the inner domain and thus gives the spherical
solution a de Sitter core. This scenario is reminiscent of the
`gravastar' picture, see \cite{MaMo2001, ViWi2003} and compare
\cite{CHLS2003}. Based on the assumption of the incompatibility of the
standard Schwarzschild spacetime with quantum mechanics, these authors
develop a setup in which this spacetime is replaced, in the
interior, by a segment of de Sitter space, separated by a shell, filled
with a `quantum fluid', from the
exterior Schwarzschild solution. These models are hoped to resolve
among other problems the black hole information paradox. Intriguingly, the gravity model proposed here
enforces the existence of a special radius at which a de Sitter core
may be merged to an approximately Schwarzschild outer domain, even without
relying on the \textit{ad hoc} introduction of any quantum fluid. Admittedly, the physical
interpretation of the positive energy solution near the special point
\( r_0 \) has to be worked out in more detail.

In the case of the negative energy analogue of the Schwarzschild solution, the
mass parameter \( m \) is negative, which makes the series coefficient \( a_{3} \)
positive (keeping \( \lambda >0 \)). Now each term of the power series solution
diverges for small radii \( r^{3}<|a_{3}| \), so the whole series does, which
has also been confirmed numerically. Hence, the singularity of the negative
energy solution does not get regularized and marks this solution as unphysical.
This is a remarkable example of how a gravity theory improves the singularity
situation by getting rid of the annoying ones while keeping those which are
important for a stable Minkowski vacuum. 

Some other points are interesting. The positive energy solutions of figure \ref{fig. Alimit}
which just regularize the metric function do only have a single horizon. The
solution regularizing also the curvature singularities has two
horizons. This confirms and exemplifies an argument of
 \cite{Hol2002}, which says that the number
of horizons in a solution of higher derivative gravity that
regularizes the curvature singularities at 
\( r=0 \) has to be even.

\begin{figure}[ht]
{\par\centering \resizebox*{0.65\columnwidth}{!}{\includegraphics{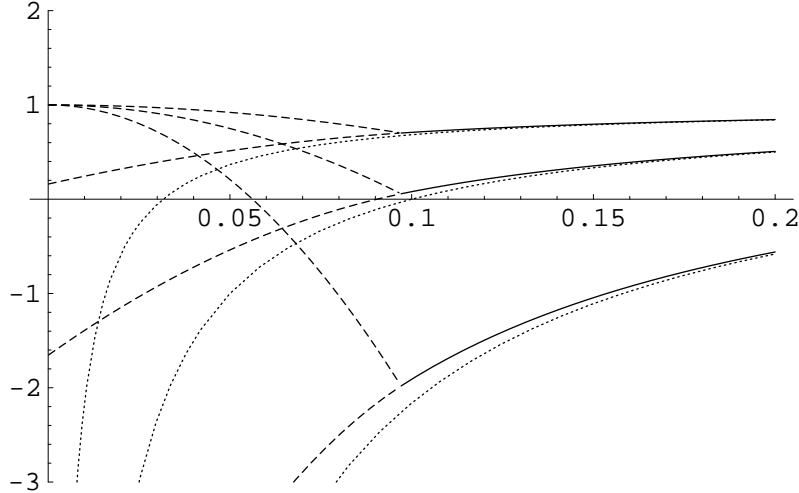}} \par}
\caption{\label{fig. baremass}\textit{The solutions of the Schwarzschild
function \protect\( A(r)\protect \) for fixed \protect\( 2m\lambda =10^{-3}\protect \),
but different ratios \protect\( 2m/\lambda =10^2,\,10,\, 1\protect \) from
bottom to top. The closely dashed lines show the corresponding Schwarzschild solutions
of general relativity with their respective horizons at \protect\( 10^{-1/2},\, 10^{-1},\, 10^{-3/2}\protect \).}}
\end{figure}

Figure \ref{fig. Alimit} has been plotted for \( 2m=1 \) and \( \lambda =10^{-3} \),
i.e., for the ratio \( 2m/\lambda \gg 1 \). The solutions undergo a qualitative
change for smaller masses. To see this, figure \ref{fig. baremass} shows the
numerical solutions for the same \( 2m\lambda  \) but for different ratios \( 2m/\lambda  \).
 For masses that are smaller than \( \lambda  \), the horizons disappear and
the Schwarzschild function stays positive for all values of \( r \). This phenomenon
has been christened a `bare mass' in \cite{Fei1998}. The existence of such
small bare masses, however, is unclear. As will be seen below, in string theory,
the parameter \( \lambda  \) would naturally be of the order of magnitude of
\( \alpha ' \). On the other hand, it is a nice feature of a gravity theory
that extremely small, but also extremely localized, energy fluctuations do only
produce a small disturbance of Minkowski spacetime without a horizon.

\section{Cancellation of ghosts\label{sec. cancel ghosts}}

Ghost freedom is a necessary requirement for any theory of gravity to be of
possible relevance in a fundamental unified quantum theory. To achieve ghost
freedom in gravity \textit{{\`a} la} Born-Infeld, one has to go beyond the basic
model presented in the preceding section. 

While keeping the nice determinantal form of the action, it is not sufficient
to consider only terms of first order in the curvature inside the determinant;
second order terms have to be included as well. The most general action, obeying
the construction principle of antisymmetrized indices used above, is of the
form
\begin{equation}
\label{eq. extmodel}
\int \sqrt{-g}\left[ \left( \det (\delta ^{A}_{B}+\lambda M^{A}_{\, \, B}+\lambda ^{2}N^{A}_{\, \, B})\right) ^{\zeta }-1\right] 
\end{equation}
where \( M_{\, \, B}^{A} \) and \( N^{A}_{\, \, B} \) contain all possible
first and second order terms, respectively. Hence, a curvature expansion of
the action directly corresponds to an expansion in orders of the parameter \( \lambda  \).
In general,\begin{subequations}\label{eq. mn}
\begin{eqnarray}
M_{\, \, B}^{A} & = & \kappa R\delta _{B}^{A}+R^{A}_{\, \, B}+\mu S^{A}_{\, \, B}\, ,\\
N_{\, \, B}^{A} & = & \left( \rho R^{2}+\sigma R_{ab}^{2}+\alpha R_{abcd}^{2}\right) \delta ^{A}_{B}+\beta RR^{A}_{\, \, B}+\tau RS^{A}_{\, \, B}+\gamma R^{A}_{\, \, C}R^{C}_{\, \, B}\nonumber \\
 &  & +\, \delta _{1}R^{A}_{\, \, C}S^{C}_{\, \, B}+\delta _{2}S^{A}_{\, \, C}R^{C}_{\, \, B}+\xi S^{A}_{\, \, C}S^{C}_{\, \, B}+\omega _{1}R_{1B}^{\, A}+\omega _{2}R_{2B}^{\, A}+\omega _{3}R_{3B}^{\, A}
\end{eqnarray}
\end{subequations}where fourteen new parameters occur which are denoted by greek
characters. The tensors \( S \) and \( R_{i} \) are defined in the appendix,
and some of their properties are given. Here, it suffices to say that \( S \)
essentially contains the Ricci tensor, and the \( R_{i} \) present different
contractions of two Riemann tensors. 

Substituting the above expressions for \( M \) and \( N \) into the curvature
expansion formula (\ref{eq. curv exp}), one finds the following linear and
quadratic terms,
\begin{equation}
\label{eq. linquad}
\frac{1}{2}\lambda f_{0}R+\frac{1}{2}\lambda ^{2}\left( f_{1}R^{2}-4f_{2}R_{ab}^{2}+f_{3}R_{abcd}^{2}\right) ,
\end{equation}
where some functions \( f_{i} \) of the parameters in the model appear. Ghost
freedom is implied by the Gauss-Bonnet combination (\ref{eq. gbcomb}) of the
quadratic curvature terms which in turn follows from the two equations \( f_{1}=f_{2}=f_{3} \).
More specifically,\begin{subequations}
\begin{eqnarray}
f_{0} & = & 1+(d-1)(d\kappa +2\mu )\, ,\\
f_{1} & = & \frac{\zeta }{4}f_{0}^{2}-\frac{d-1}{2}(d\kappa ^{2}+4\kappa \mu -2d\rho -4\tau )-\kappa -\mu ^{2}+\beta +2\xi \, ,\\
f_{2} & = & -\frac{d(d-1)}{4}\sigma -\frac{1}{2}(\delta _{1}+\delta _{2}-\mu +\omega _{3})+\frac{d-2}{4}(\mu ^{2}-2\xi )\, ,\\
f_{3} & = & d(d-1)\alpha +\frac{1}{2}\gamma -\frac{1}{4}+\omega _{1}+2(d-1)\omega _{2}-\omega _{3}\, .
\end{eqnarray}
\end{subequations}This implies that there is a huge class of ghost-free models;
this class is a thirteen-parameter family because the fourteen new parameters
are only constrained by the following two equations,\begin{subequations}\label{eq. gaussbonnet}
\begin{eqnarray}
 & d(d-1)(\sigma +4\alpha )+2(\gamma +\delta _{1}+\delta _{2})-(d-2)(\mu ^{2}-2\xi ) & \nonumber \\
 & -\, 2\mu +4\omega _{1}+8(d-1)\omega _{2}-2\omega _{3}-1\, =\, 0\, , & \\
 & \zeta \left( 1+(d-1)(d\kappa +2\mu )\right) ^{2}-(d-1)(2d\kappa ^{2}+8\kappa \mu -4d\rho -d\sigma -8\tau ) & \nonumber \\
 & +\, 2(\delta _{1}+\delta _{2}-\mu +\omega _{3}-2\kappa +2\beta )-(d+2)(\mu ^{2}-2\xi )\, =0\, , & 
\end{eqnarray}
 \end{subequations}and the expansion parameter \( \lambda  \) is free. 

Unfortunately, the equations of motion for these models, even in the simple
case of a spherically symmetric spacetime, are extremely involved. It is expected
that, at least in some cases, the nice features, and essentially the regularization
property, of the basic model of the preceding section are kept. This has been
checked, for instance, for the specific parameter values \( \lambda \neq 0 \),
\( \beta =-\frac{1}{24} \) and \( \gamma =\frac{1}{2} \) (with all remaining
parameters vanishing). All properties were found to be qualitatively similar.

Clearly, if the gravity theories in Born-Infeld form were expected to play an
important role in any fundamental theory, one would need further physical principles
to narrow down the wide range of possible choices. One such, very constraining,
principle might be the requirement of supersymmetrizability of the gravitational
action. Another might be the idea that such an action could be derived as an
effective action from string theory, as it has been possible in the case of
Born-Infeld electrodynamics.

\section{Correspondence to string theory\label{sec. string theory}}

Two different methods have been used to deduce the Born-Infeld action from string
theory. One uses the Polyakov path integral \cite{FrTs1985}, in which the open
string degrees of freedom are integrated out, in order to derive the effective
action for the appropriate background fields. The other method requires conformal
invariance on the string worldsheet and constructs the corresponding beta functions
\cite{ACNY1987}. The conditions that they vanish can be obtained as equations
of motion from the effective action. The first approach is not available in
the gravitational case. The Polyakov path integral is the connected generating
functional and contains the effect of massless poles, and so it should not have
a meaningful low energy expansion \cite{GKP1986}. The approach via the gravitational
beta functions is a valid one. This has been shown, for example, in
\cite{HuTo1988} for general string gravity backgrounds, including the
antisymmetric tensor field and the dilaton. 
But so far, this method has not been as successful as in
the Born-Infeld case. The effective gravity action derived from string theory
is only known for some low orders in the curvature.

Nevertheless, the known first few orders of effective gravity deduced from string
theory can be used to narrow down the choice of parameters for the ghost-free
models of the preceding section. This is done under the working hypothesis that
such an action might indeed be of importance in a fundamental theory. A ghost-free
model has to be taken because string induced gravity is ghost-free \cite{JJL1988}.
The effective action employed here contains curvature terms up to the third
order \cite{Gra1987} and reads
\begin{eqnarray}
\int \sqrt{-g}\left[ R+\alpha '\textrm{Tr}(R\cdot R)+\alpha '^{2}\left( \frac{1}{2}\textrm{Tr}(R\cdot R\cdot R)-\frac{1}{12}\textrm{Tr}(R\cdot R_{3})\right. \right. \qquad \qquad  &  & \nonumber \\
\left. \left. +\, \frac{3}{2}\textrm{Tr}(S\cdot R\cdot R)-\frac{3}{8}R\textrm{Tr}(R\cdot R)\right) \right]  &  & 
\end{eqnarray}
where the notation introduced in the appendix has been used. For
illustrational purposes, and because the Born-Infeld style gravities at
present are purely bosonic theories, the action for a bosonic \( \sigma \)-model
is used in this section. In the supersymmetric case, the situation
presents itself in a different light. Then the \( \sigma \)-models do
not lead to third order curvature corrections of the Einstein
action \cite{MeTs1987}. Terms of fourth order, however, are important
\cite{GrWi1986} and have been paid a lot of attention. This is also
due to the fact that they are the first non-trivial,
supersymmetrizable (see, e.g., \cite{BeRo1989,RSW1992}) corrections arising from
string theories which are compactified to four dimensions. The second order
terms appear in the Gauss-Bonnet combination, while the third order
ones are not supersymmetrizable.

Note that the above
action does seem to contain ghosts. This is because string scattering amplitudes,
and beta functions, are calculated for external gravitons that are on Einstein
shell, i.e., that satisfy the (linearized) vacuum Einstein equations \( R_{\mu \nu }=0 \).
This means that, in second order of the curvature, the terms \( R^{2} \) and
\( R_{ab}^{2} \) simply can be added with the appropriate factors to achieve
the Gauss-Bonnet combination \cite{DeRe1986} (the same can be effected 
by means of appropriate field redefinitions \cite{HuTo1988}). More generally, it means that
a string-induced effective gravity action will never contain any terms that
involve at least twice the Ricci tensor or Ricci scalar. Upon variation, one
of them will always remain so that these terms vanish on shell,
compare \cite{JJM1989}. So the first
allowed simplification in the Born-Infeld gravity models, which does not interfere
with their consistency with string theory, is to set \( \rho =\sigma =\tau =\xi =0 \);
compare equation (\ref{eq. mn}).

In addition to the two equations (\ref{eq. gaussbonnet}) that enforce Gauss-Bonnet
on the quadratic level, there are further equations that constrain the remaining
parameters. The first comes from comparing the first two orders of the curvature,
see (\ref{eq. linquad}ff), and the string-induced action,
\begin{equation}
\label{eq. consistency1}
\alpha 'f_{0}=4\lambda f_{3}\, .
\end{equation}
Four equations are derived from comparing the third order terms in the curvature
expansion. For this purpose, substitute (\ref{eq. mn}) into (\ref{eq. curv exp})
and utilize the trace relations (\ref{eq. trace1})-(\ref{eq. trace4}), which
leads to\begin{subequations}\label{eq. consistency2}
\begin{eqnarray}
\alpha '^{2}f_{0} & = & -\frac{4}{3}\lambda ^{2}\left( 3\gamma -1\right) ,\\
\alpha '^{2}f_{0} & = & 24\lambda ^{2}\omega _{3}\, ,\\
\alpha '^{2}f_{0} & = & -\frac{4}{3}\lambda ^{2}\left[ \gamma \mu -\mu +\delta _{1}+\delta _{2}+2\omega _{2}+2\mu \left( \omega _{1}+(d-2)\omega _{2}-\omega _{3}\right) \right] ,\\
\alpha '^{2}f_{0} & = & -\frac{4}{3}\lambda ^{2}\left[ f_{0}\left\{ 4\zeta \left( \omega _{1}+2(d-1)\omega _{2}-\omega _{3}+d(d-1)\alpha \right) -8\alpha +\zeta (2\gamma -1)\right\} \right. \nonumber \\
 &  & \qquad \qquad \quad \, \left. -\, 4\left\{ \kappa (\gamma -1)+\beta +2\kappa \left( \omega _{1}+2(d-1)\omega _{2}-\omega _{3}\right) +4\mu \omega _{2}\right\} \right] .
\end{eqnarray}
\end{subequations}A last equation in the comparison at third order arises from
the combination \( \textrm{Tr}(R\cdot R_{1}) \) that is not present in the
string-induced action. It appears, however, in the curvature expansion of the
Born-Infeld style models, and so it has to cancel. This simply gives the condition
\begin{equation}
\omega _{1}=0\, .
\end{equation}

So all extended Born-Infeld gravity models of the kind defined in section \ref{sec. cancel ghosts}
in (\ref{eq. extmodel}ff) whose parameters satisfy the constraint equations
(\ref{eq. gaussbonnet}) and (\ref{eq. consistency1})-(\ref{eq. consistency2})
are consistent with string-induced effective gravity at least up to third order
of the curvature. In principle, this check could be continued to higher orders
with the aim of finding further consistency conditions (or possibly an inconsistency
with the theoretical \( \sigma \)-model predictions). This is, however, not a very interesting
task. More important is the following observation.

From the constraint equations, it is clear that the expansion parameter \( \lambda  \)
is proportional to the inverse string tension, \( \lambda \sim \alpha ' \).
This, again, is very reminiscent of Born-Infeld electrodynamics, where the action
takes into account all orders in \( \alpha ' \) and, hence, all orders of the
field strength tensor, which is responsible for the regularization of the electric
field of point charges. The fascinating idea here is the suggestion of a very
similar mechanism for gravity. By taking into account all orders of \( \alpha ' \)
in curvature corrections in the string-induced gravity action, string theory
might resolve the singularity problems of general relativity. The theories presented
here are the first examples of gravity theories doing just that. Whether or
not one of them is indeed derivable from string/M-theory cannot be answered
at present.

\section{Discussion\label{sec. discussion}}

A simple recipe has been presented for constructing gravity theories with an
action in determinantal form, very similar to the action of Born-Infeld electrodynamics.
The construction is based on the observation that the Schwarzschild singularity
can only be recognized from the second order curvature invariant \( R_{abcd}^{2} \),
but not from those invariants constructed from the Ricci tensor. Hence, a gravity
theory that expects to cure singularities, by placing upper bounds on \( R_{abcd}^{2} \),
or otherwise, should include the full Riemann tensor. The actual formulation
of the new class of theories presented in this article utilizes the symmetries
of the Riemann tensor, considering only tensors with pairs of antisymmetrized
indices. 

For the simplest version of Born-Infeld style gravity, the analogue of the spherically
symmetric Schwarzschild solution has been analyzed. A different behaviour has
been found in different mass regimes. In all cases the Schwarzschild singularity,
at least for the positive energy solution, may be removed (a detailed
discussion is found in section \ref{sec. sing}). For large masses, compared
to some expansion parameter, the horizon remains; small masses, on the other
hand, become `bare masses' without a horizon. The negative energy Schwarzschild
solution remains singular and is thus distinguished to be unphysical, which
is necessary in order to have a stable Minkowski vacuum configuration. The bare
masses are a nice feature in the sense that extremely small yet localized energy
fluctuations would only produce small spacetime distortions, without generating
horizons and thereby changing the topology. These characteristics make the new
Born-Infeld style gravities nice examples of theories, in which the annoying
singularities are removed, whereas the important ones remain.

It has been shown that extended models can be defined in which the ghosts, in
a perturbative expansion around flat Minkowski spacetime, cancel. The improved
behaviour of these theories with respect to general relativity is kept, at least
for some models. Unfortunately, the calculations, even for the simple spherically
symmetric spacetime, are quite involved, so that it remains unclear whether
all extended, ghost-free models have improved singularity properties. 

Ghost-freedom is essential, if the Born-Infeld gravities were expected to be
derivable from an underlying, unified theory such as string or
M-theory. The consistency
with string-induced gravity, has been illustrated for a bosonic \(
\sigma \)-model, at least up to third order in the curvature. In checking this correspondence, all influence of the dilaton field
and of the antisymmetric tensor field of the string gravity backgrounds has
been neglected. It might be of interest to pursue extensions of the Born-Infeld
style gravities which try to incorporate these fields. The value of the string
theory against gravity comparison is not so much a proof, showing exactly what
kind of gravity is induced by string theory, but much more the exhibition of
the following interesting and important mechanism. 

A way has been suggested in which string theory, by inclusion of all orders of \( \alpha ' \)
corrections, corresponding to all orders of the curvature in the string-induced
gravity action, might actually regularize gravitational singularities. The final
gravity theory induced from string or M-theory thus is expected to have at least
all the good features of the gravities \textit{{\`a} la} Born-Infeld,
which are nice examples realizing this idea. Furthermore,
it would probably be supersymmetrizable, and it would be interesting to investigate
whether any of the models here has this property. 



\appendix

\section{Some technicalities}

This appendix contains some calculations and definitions needed in the main
text, especially in section \ref{sec. cancel ghosts}, which considers the ghost-free
models of Born-Infeld gravity, and in section \ref{sec. string theory}, where
these models are compared to string-induced effective gravity. 

The following tensors are needed for the generalized models. They contain the
Ricci tensor at the first order level (which cannot be obtained by tracing \( R^{A}_{\, \, B} \))
and some special combinations of two Riemann tensors in second order of the
curvature:\begin{subequations}
\begin{eqnarray}
S^{A}_{\, \, B} & \equiv  & R_{\, \, b_{1}}^{a_{1}}\delta _{b_{2}}^{a_{2}}-R^{a_{2}}_{\, \, b_{1}}\delta ^{a_{1}}_{b_{2}}-(b_{1}\leftrightarrow b_{2})\, ,\\
R^{\, A}_{1B} & \equiv  & R^{a_{1}a_{2}cd}\left( R_{cb_{1}db_{2}}-R_{cb_{2}db_{1}}\right) ,\\
R^{\, A}_{2B} & \equiv  & R^{a_{1}cde}R_{b_{1}cde}\delta ^{a_{2}}_{b_{2}}-R^{a_{2}cde}R_{b_{1}cde}\delta ^{a_{1}}_{b_{2}}-(b_{1}\leftrightarrow b_{2})\, ,\\
R^{\, A}_{3B} & \equiv  & R^{a_{1}cd}_{\quad \, \, \, b_{1}}R^{a_{2}}_{\, \, \, \, cdb_{2}}-R^{a_{2}cd}_{\quad \, \, \, b_{1}}R^{a_{1}}_{\, \, \, \, cdb_{2}}-(b_{1}\leftrightarrow b_{2})\, .
\end{eqnarray}
\end{subequations}These tensors and their products, including also the Riemann
tensor \( R^{A}_{\, \, B} \), obey the following important trace relations,
in which the notation \( (R\cdot S)^{A}_{\, \, B}=R^{A}_{\, \, C}S^{C}_{\, \, B} \)
is used,\begin{subequations}\label{eq. trace1}
\begin{eqnarray}
\displaystyle \textrm{Tr}\, \delta =\frac{d(d-1)}{2}\, ,\, \, \quad  & \textrm{Tr}\, R={\frac{1}{2}}R\, , & \, \, \quad \textrm{Tr}\, S=(d-1)R\, ,\\
\textrm{Tr}\, R_{1}=\frac{1}{2}R_{abcd}^{2}\, ,\, \, \quad  & \textrm{Tr}\, R_{2}=(d-1)R_{abcd}^{2}\, , & \, \, \quad \textrm{Tr}\, R_{3}=R_{ab}^{2}-\frac{1}{2}R_{abcd}^{2}\, ,
\end{eqnarray}
\end{subequations}and\begin{subequations}
\begin{eqnarray}
\textrm{Tr}(R\cdot R) & = & \frac{1}{4}R_{abcd}^{2}\, ,\\
\textrm{Tr}(R\cdot S) & = & R_{ab}^{2}\, ,\\
\textrm{Tr}(R\cdot R_{1}) & = & \frac{1}{2}R^{abcd}R_{cdef}R^{e\, \, f}_{\, \, a\, \, b}\, ,\\
\textrm{Tr}(R\cdot R_{2}) & = & 2\textrm{Tr}(SRR)\, ,\\
\textrm{Tr}(R\cdot R_{3}) & = & R^{abcd}R_{cefa}R^{\, \, \, ef}_{d\, \, \, \, \, b}\, ,
\end{eqnarray}
\end{subequations}as well as\begin{subequations}
\begin{eqnarray}
\textrm{Tr}(S\cdot S) & = & R^{2}+(d-2)R_{ab}^{2}\, ,\\
\textrm{Tr}(S\cdot R_{1}) & = & 2\textrm{Tr}(S\cdot R\cdot R)\, ,\\
\textrm{Tr}(S\cdot R_{2}) & = & 2(d-2)\textrm{Tr}(S\cdot R\cdot R)+4R\textrm{Tr}(R\cdot R)\, ,\\
\textrm{Tr}(S\cdot R_{3}) & = & -2R_{ab}R_{cd}R^{acdb}-2\textrm{Tr}(S\cdot R\cdot R)\, ,
\end{eqnarray}
\end{subequations}and\begin{subequations}\label{eq. trace4}
\begin{eqnarray}
\textrm{Tr}(R\cdot R\cdot R) & = & \frac{1}{8}R^{abcd}R_{cdef}R^{ef}_{\, \, \, \, \, \, ab}\, ,\\
\textrm{Tr}(S\cdot R\cdot R) & = & R_{ab}R^{acde}R^{b}_{\, \, \, cde}\, .
\end{eqnarray}
\end{subequations}In all calculations involving tensors with antisymmetrized
index pairs, care has to be taken that a sum over capital indices is equivalent
to a sum over unordered normal indices multiplied by a factor of one half, i.e.,
symbolically
\begin{equation}
\sum _{A}\, \longleftrightarrow \, \frac{1}{2}\sum _{a_{1},\, a_{2}}\, .
\end{equation}

The trace identities are necessary in order to evaluate the curvature expansion
of the Lagrangian, i.e., of the function
\begin{equation}
\displaystyle \left( \det (\delta ^{A}_{B}+\lambda M^{A}_{\, \, B}+\lambda ^{2}N^{A}_{\, \, B})\right) ^{\zeta }
\end{equation}
where \( M \) and \( N \) are of first or second order in the curvature, respectively.
This curvature expansion is equivalent to an expansion around \( \lambda =0 \)
and is obtained by using the matrix formula
\begin{equation}
\det (\delta +C)=1+\sum _{m=1}^{\infty }\frac{1}{m!}\left( \sum _{n=1}^{\infty }\frac{(-1)^{n+1}}{n}\textrm{Tr}(C^{n})\right) ^{m}
\end{equation}
which follows from the identity \( \ln \det (\delta +C)=\textrm{Tr}\ln (\delta +C) \)
and the series expansions of the logarithm and the exponential. Up to third
order, the result is
\begin{eqnarray}
 &  & 1+\zeta \left[ \lambda \textrm{Tr}M+\lambda ^{2}/2\left( 2\textrm{Tr}N+\zeta (\textrm{Tr}M)^{2}-\textrm{Tr}(M^{2})\right) +\lambda ^{3}/6\left( \zeta ^{2}(\textrm{Tr}M)^{3}+6\zeta \textrm{Tr}M\textrm{Tr}N\right. \right. \nonumber \label{eq. curv exp} \\
 &  & \qquad \qquad \qquad \left. \left. -\, 3\zeta \textrm{Tr}M\textrm{Tr}(M^{2})-6\textrm{Tr}(MN)+2\textrm{Tr}(M^{3})\right) +\mathcal{O}(\lambda ^{4})\right] .\label{eq. curv exp} 
\end{eqnarray}


\section{Corrigendum}
Class. Quantum Grav. \textbf{21} (2004) 5297.\\

In the paper above, a method for the construction of novel gravity
actions of Born-Infeld type is presented. (Complete
information on the Riemann tensor is encoded and
could enable improved singularity behaviour, which is embellished by 
relations to theories with tidal acceleration bounds \cite{ScWo04}.) 

A few points need correction and clarification. Expression {(11)}
for the effective action of the basic model on static spherically
symmetric spacetimes is incorrect; it should read 
\setcounter{equation}{10}
\begin{eqnarray}
& &\hspace{-2cm}\int dr \sqrt{AB}r^2
  \left[\frac{1}{2r}A^{-2/3}B^{-3/2}\Big((\lambda+B(r^2-\lambda))(2rAB+\lambda 
  A')^2(2rB^2-\lambda
  B')^2\times\right.\nonumber\\
& & \qquad\qquad\left.\times(4A^2B^2-\lambda(AA'B'+{A'}^2B-2AA''B))\Big)^{1/6}-1\right].
\end{eqnarray}
This correction affects the term $-1$ within the square brackets
of the original equation in the paper which is replaced by
$-(AB)^{-1/2}r$ here. The curvature invariants on four-dimensional de
Sitter space, given below {figure 1}, should read $R=12/\lambda$,
$R_{ab}^2=36/\lambda^2$ and $R_{abcd}^2=24/\lambda^2$, in the usual
convention that the Riemann tensor of a space of constant curvature
$1/\lambda$ is
$R_{abcd}=\left(g_{ac}g_{bd}-g_{ad}g_{bc}\right)/\lambda$. The first
misprint was pointed out by Deser \textit{et al.}
\cite{Deser}; subsequent computations are unaffected.    
 
Another important point has been raised by the authors of
\cite{Deser}. Given any physical theory in terms of its action, it is
generally not the case that solutions to the equations of motion of an
effective theory obtained by a truncation of the original action (by
means of a specific ansatz for the relevant fields) are also solutions
to the equations of motion of the original theory. In other words,
truncations in this sense and variation generally do not
commute. Nevertheless, it has been shown \cite{Pal1979} that the
effective action {(11)} for a static spherically symmetric spacetime
ansatz with two undetermined functions $A$ and $B$ of the radial
coordinate may be considered. 

In the paper, a further truncation of
the static spherically symmetric ansatz, setting $B=1/A$, has been
used in deriving the effective action (13) and the equation of motion
(14). This truncation is not related to any symmetry; thus, equation
(14) is necessary, but not sufficient: any solution of the
equations for $A$ and $B$ from (11) which satisfies $B=1/A$ solves
(14), but the converse is not generally true. Therefore, the validity of any
solution of (14) has to be checked by substituting it into at least
one of the equations for $A$ or $B$ from (11). The procedure of using the
simpler equation (14) should be considered merely a means of
suggesting solutions for the original equations.   

The paper gives $c_1r^3,\,c_1r^2$ and $c_1r+c_2$ for arbitrary
constants $c_i$ as solutions of (14). Unfortunately, a check of the validity of these solutions in the equations derived from
(11) was omitted, and we will do this check now. We find that $c_1r^3$ is
invalid. However, $c_1r^2$ gives valid solutions, provided either
$c_1=-3$ or $c_1=1$, where the constant $c_1$ is determined as a real
solution of the equation $2|c_1|=c_1+c_1^2$ (which implies the
equations from (11) with $B=1/A$). 
For the linear expression $c_1r+c_2$, a denominator vanishes
in the equations derived from both actions (11) and (13), so that this
expression cannot be a stationary solution. 


It has not been properly checked yet whether the previously given
numerical solutions of (14) are correct. Although a substitution of
the solution into the equations from (11) produces `small' errors, and
although the error associated with the equation of motion for $A$
oscillates around zero, such a check is not conclusive in many
examples of differential equations. A proper check would involve the
integration of both equations for $A$ and $B$.
  


\acknowledgments

It is a pleasure to thank Paul K. Townsend for helpful discussions, and Bob Holdom, Donald Marolf and Frederic P. Schuller for useful e-mails. 
I also wish to thank Stanley Deser, Bayram Tekin and Joel Franklin for e-mail
correspondence, drawing to my attention refs. \cite{Pal1979,FeTo2002} on symmetry reductions and pointing out some mistakes. 
Financial support from the Gates Cambridge Trust is gratefully acknowledged.



\end{document}